\documentclass[aps,prl,reprint,groupedaddress]{revtex4-1}

\usepackage{graphicx}
\newcommand{\GeV}{\,\mbox{GeV}}
\newcommand{\as}{\alpha_s}
\newcommand{\mupi}{\mu_\pi^2}
\newcommand{\mug}{\mu_G^2}
\newcommand{\rd}{\rho_D^3}
\newcommand{\rls}{\rho_{LS}^3}

\long\def\symbolfootnote[#1]#2{\begingroup%
\def\thefootnote{\fnsymbol{footnote}}\footnote[#1]{#2}\endgroup}

\def \be{\begin{equation}}
\def \ee{\end{equation}}
\newcommand{\bea}{\begin{eqnarray}}
\newcommand{\eea}{\end{eqnarray}}
\def \nn{\nonumber}

\begin{document}

\title{Precision determination of the CKM element $V_{cb}$}

\author{Andrea Alberti}
\author{Paolo Gambino}
\author{Kristopher J. Healey}

\affiliation{Dipartimento di Fisica, Universit\`a di Torino \& INFN, Torino, 10125 Torino, Italy}
\author{Soumitra Nandi}
\affiliation{Physics Department, Indian Institute of Technology, Guwahati, 781 039, India}


\begin{abstract}
We extract the magnitude of the CKM matrix element $V_{cb}$ and the most relevant  parameters of the Heavy Quark Expansion from data of inclusive semileptonic $B$ decays. Our calculation includes the recently computed $O(\alpha_s
\Lambda_{QCD}^2/m_b^2)$ corrections and
a careful  estimate of the residual theoretical uncertainty. Using a recent determination of the charm quark mass, we obtain $|V_{cb}|= (42.21\pm 0.78)\times 10^{-3}$ and $m_b^{kin}(1\rm GeV)=(4.553\pm 0.020)$\,GeV.

\end{abstract}

\pacs{13.25.Hw, 14.40.Nd, 12.15.Hh, 12.15.Ff}

\maketitle

\section{Introduction}
The Cabibbo-Kobayashi-Maskawa (CKM) mechanism  of quark flavor violation is one of the main components of the Standard Model (SM) of fundamental interactions \cite{Cabibbo:1963yz,Kobayashi:1973fv}. It accommodates very well all of the observed CP violation, as well as the flavor changing phenomena studied  at kaon experiments, the B factories, and at high-energy colliders like the LHC, see  \cite{Antonelli:2009ws,Bevan:2014iga} for recent reviews.
The $ 3\times 3$ unitary CKM matrix, which  parameterizes  flavor violation in this context, has only four independent parameters.
 While they are strongly constrained by present data,  
any improvement would be welcome as it would sharpen our tools for future tests of the SM.

In particular, more precise measurements of $|V_{cb}|$, the CKM element controlling charged current $b\leftrightarrow c$ transitions,  would  crucially help the 
search for new physics  in rare decays, which requires accurate SM predictions.
   Indeed, the present $\sim 2\%$ error on this single CKM element represents   the dominant uncertainty on the SM prediction of important flavor-changing neutral current decays such as $B_s\to \mu^+\mu^-$ 
   \cite{Bobeth:2013uxa}, $K^+\to \pi^+\nu\bar \nu$, 
   and  $K_L\to \pi^0\nu\bar \nu$ 
   \cite{Brod:2010hi}, as well as of the  CP violation parameter $\varepsilon_K$ \cite{Bailey:2014qda}.

Direct information on $|V_{cb}|$ can be obtained from inclusive and exclusive semileptonic $B$ decays to charmed hadrons, which are subject to different theoretical and experimental systematics.
In the first case, the Operator Product Expansion (OPE) allows us to 
describe the relevant non-perturbative physics in terms of a small number of  parameters that can be extracted from experiment. In the case of the  exclusive decays 
$B\to D^{(*)}\ell \bar \nu$,  the form factors have to be computed by non-perturbative methods, {\it e.g.}\  lattice QCD.
The most precise recent results of each method are
\be
|V_{cb}|= (42.42\pm 0.86)\times 10^{-3}\label{incl}
\ee
from a global fit to  inclusive semileptonic moments \cite{Gambino:2013rza}, and
\be
|V_{cb}|= (39.04\pm 0.49_{exp}\pm 0.53_{lat}\pm 0.19_{ QED})\times 10^{-3}\label{excl}
\ee
from an unquenched lattice QCD calculation of the zero recoil form factor of $B\to D^* l \nu$ by the Fermilab-MILC collaboration \cite{Bailey:2014tva}.  
They disagree by $3\sigma$, which remains a long-standing tension.
There exist also less precise determinations of $|V_{cb}|$ based on heavy quark sum rules and the decay $B\to D l \nu$, see \cite{Bevan:2014iga} for a review.

It is also possible to determine $|V_{cb}|$ indirectly, using the CKM unitarity relations together with CP violation and flavor data, excluding the above direct information: SM analyses  
by the UTfit and CKMFitter collaborations give $(42.05\pm 0.65)\times 10^{-3}$ \cite{Bona:2006ah} and $(41.4^{+2.4}_{-1.4})\times 10^{-3}$  \footnote{CKMfitter Group (J. Charles {\it et al.}), Eur. Phys. J. C41, 1 (2005) [hep-ph/0406184], updated results and plots available at: http://ckmfitter.in2p3.fr},
    both closer to the inclusive value of Eq.\,(\ref{incl}).

In principle, the lingering discrepancy between the values of $|V_{cb}|$ extracted from inclusive decays and from $B\to D^* l \nu$ could be ascribed to physics beyond the SM, 
as the $B\to D^*$ transition is sensitive only to the axial-vector component of the $V-A$ charged weak current. However, the new physics effect should be sizable (8\%),
and 
would require new interactions ruled out by electroweak  constraints on the effective $Zb\bar b$ vertex \cite{Crivellin:2014zpa}.
The most likely explanation of the discrepancy between Eqs.\,(\ref{incl},\ref{excl}) is therefore a problem in the theoretical or experimental analyses of semileptonic decays.

In this Letter we  focus on  the inclusive extraction of $|V_{cb}|$,  
including all contributions of $O(\alpha_s \Lambda_{\rm QCD}^2/m_b^2)$, whose calculation has been recently completed \cite{Becher:2007tk,Alberti:2012dn,Alberti:2013kxa},   and discuss how this improvement affects the results.

\section{The calculation}
Let us briefly review the calculation of the quantities that enter the 
inclusive analysis. The OPE allows us to write sufficiently inclusive quantities (typically the width and the first few moments of kinematic distributions)
as double series in
$\alpha_s$ and $\Lambda_{QCD}/m_b$. The expansion in powers of the heavy quark mass starts at $O(1/m_b^2)$ \cite{Chay:1990da,Bigi:1992su, Blok:1993va,Manohar:1993qn} and involves
the $B$-meson expectation values of local operators of growing dimension. These non-perturbative parameters can be constrained from the measured values of the normalized 
moments 
of the lepton energy and invariant hadronic mass  distributions in  $B\to X_c\ell\nu$ decays:
\bea
  \langle
  E^n_\ell\rangle&=&\frac{1}{\Gamma_{E_\ell>E_\mathrm{cut}}}\int_{E_\ell>E_\mathrm{cut}}
   E^n_\ell\ \frac{d\Gamma}{dE_\ell} \ dE_\ell~, \nonumber\\
     \langle
  m^{2n}_X\rangle&=&\frac{1}{\Gamma_{E_\ell>E_\mathrm{cut}}}\int_{E_\ell>E_\mathrm{cut}}
  m^{2n}_X\ \frac{d\Gamma}{dm^2_X}\ dm^2_X~.\nonumber
\eea
where $E_\ell$ is the lepton energy,   $m_X^2$ the invariant hadronic squared mass, and 
$E_\mathrm{cut}$ an experimental threshold on the lepton energy applied by some of the experiments.
Since the physical information of moments of the same type is highly correlated, 
for $n>1$  it is better to employ {\it central} moments, computed relative to 
$\langle E_\ell\rangle$ and $\langle m^2_X\rangle$.
The information on the non-perturbative parameters obtained from a fit to the moments enables us to extract  $|V_{cb}|$ from the total semileptonic width \cite{Gambino:2004qm,Bauer:2004ve,Buchmuller:2005zv}.
\begin{table}
  \begin{center} \begin{tabular}{lllllll}
    \hline 
$a^{(1)}$& $a^{(2,\beta_0)}$& $a^{(2)}$  & $p^{(1)}$ &$g^{(0)}$ & $g^{(1)} $& $d^{(0)}$ 
 \\ \hline
   -0.95 & -0.47 & 0.71&  0.99  &-1.91 & -3.51 &-16.6 
   \\    
 -1.66 & -0.43 & -2.04&  1.35  &-1.84 & -2.98 &-17.5 
 \\
   -1.24 & -0.28 & 0.01&  1.14  &-1.91 & -3.23 &-16.6\\ 
    \hline 
  \end{tabular} \end{center}
    \caption{ \label{tab:1} Coefficients of (\ref{width}) for $m_b^{kin}(1\rm GeV)=4.55$GeV
    and with the
    charm mass  in the kinetic scheme, $m_c^{kin}(1\rm GeV)=1.091$GeV (first row),
    and in the  $\overline{\rm \small MS}$ scheme, $\overline{m}_c(3\rm GeV)=0.986$GeV (2nd row) and $\overline{m}_c(2\rm GeV)=1.091$GeV (3rd row).
         }
\end{table}

The expansion for the total semileptonic width is
\bea
\Gamma_{sl} &=&\! \Gamma_0 \Big[ 1+a^{(1)} \frac{\as(m_b)}{\pi}\! +a^{(2,\beta_0)}\beta_0\!\left(\frac{\as}{\pi}\right)^2 \!\!+a^{(2)}\!\!\left(\frac{\as}{\pi}\right)^2 \!  \nn
\\ &&\!+
\left(\!-\frac12  +p^{(1)}  \, \frac{\as}{\pi}\right)\nonumber
 \frac{\mu_\pi^2}{m_b^2}   + \left(g^{(0)} +g^{(1)}\, \frac{\as}{\pi}\right)\frac{\mu_G^2(m_b)}{m_b^2}  \\
&& \left. \ \ \ \ \ \ \
+
 d^{(0)} \frac{\rho_D^3}{m_b^3} -g^{(0)} \frac{\rho_{LS}^3}{m_b^3} + {\rm higher\  orders}
  \right], \label{width}
\eea
where $\Gamma_0= A_{ew}|V_{cb}^2| G_F^2 m_b^5 (1\!-\!8\rho\!+\!8\rho^3\! -\!\rho^4 \!-\!12\rho^2 \ln\rho)/192 \pi^3$ is the tree level free quark decay width, $\rho=m_c^2/m_b^2$, and $A_{ew}= 1.014$  the leading electroweak correction. We have split the 
  $\as^2$ coefficient into a BLM piece proportional to  $\beta_0=9$ (with three massless active quark flavors) and a remainder. 
The expansions for the moments have the same structure. 
The parameters $\mupi, \mug, \rd,\rls$ are the $B$ meson expectation values of the relevant 
dimension 5 and 6 local operators.

In Eq.\,(\ref{width}) and in the calculation of all the moments we have included the complete one  and two-loop perturbative corrections \cite{Trott:2004xc,Aquila:2005hq,Pak:2008qt,              
Melnikov:2008qs,Biswas:2009rb,Gambino:2011cq}, as well as $1/m_b^{2,3}$ power corrections
 \cite{Bigi:1992su, Blok:1993va,Manohar:1993qn,Gremm:1996df}.
We neglect contributions of
order $1/m_b^4$ and $1/m_Q^5$ \cite{Mannel:2010wj}, which appear to lead to a very small shift in $|V_{cb}|$,
but we include for the first time the perturbative corrections to the leading power suppressed contributions \cite{Becher:2007tk,Alberti:2012dn,Alberti:2013kxa} to the width 
(see also \cite{Mannel:2014xza} for the limit $m_c\to 0$)
and to all the moments \footnote{The calculation in the presence of experimental cuts is not trivial; details will be given in a future publication.}.

  The coefficients $a^{(i)},g^{(i)},p^{(1)},d^{(0)}$ in Eq.\,(\ref{width}) are 
  functions of $\rho$ and of various unphysical scales, such as the one of $\as$. 
They are given in Table 1 for specific values of the quark masses.
We use the kinetic scheme \cite{Bigi:1996si} with cutoff at 1\,GeV for $m_b$ and the OPE parameters and three different options for the charm mass.


\section{The global fit}
The available measurements of the semileptonic moments \cite{Bevan:2014iga} and the 
recent, precise determinations of the heavy quark masses significantly constrain
 the parameters entering Eq.\,(\ref{width}), making  possible 
a determination of $|V_{cb}|$ whose uncertainty is dominated by our ignorance of higher 
order effects. Duality violation effects can be constrained {\it a posteriori}, by checking 
whether the OPE predictions fit the experimental data, but this again depends on precise 
OPE predictions.

We perform a fit to the semileptonic data listed in Table 1 of Ref.\,\cite{Gambino:2013rza} 
with $\as(4.6\rm GeV)=0.22$
and employ a few additional inputs. Since the moments are mostly sensitive to $\approx m_b-0.8 \,m_c$, it is essential to include information on at least one of the heavy quark masses. Because of its smaller absolute uncertainty, $m_c$ is preferable. Among
recent $m_c$ determinations \cite{Chetyrkin:2009fv,Dehnadi:2011gc,Chakraborty:2014aca} we choose  $\overline{m}_c(3\rm GeV) =0.986(13)$GeV  \cite{Chetyrkin:2009fv}, 
although we will discuss the inclusion of $m_b$ determinations as well.
We also include a loose bound on the chromomagnetic expectation value from the 
$B$ hyperfine splitting, $\mug(4.6{\rm GeV})=0.35(7)$GeV$^2$.
Finally, as all observables depend very weakly on $\rls$, we use the heavy quark sum rule constraint $\rls=-0.15(10)$GeV$^3$.
\begin{table}[t]
  \begin{center} \begin{tabular}{cccccccc}\hline
 $m_b^{kin}$ & $\overline m_c(3\GeV)$   &  $\mupi $ &$\rd$ &$\mug$ & $\rls$  & ${\rm BR}_{c\ell\nu}$ & $10^3|V_{cb}|$ \\ \hline
     4.553 &0.987 & 0.465 & 0.170 & 0.332 & -0.150 & 10.65 & 42.21 \\  
    0.020 & 0.013 & 0.068 & 0.038 & 0.062 & 0.096 & \ 0.16 & \ 0.78\\ \hline
   1
 & 0.508 & -0.099 & 0.142 & 0.596 & -0.173 & -0.075 & -0.418\\
  & 1 & -0.013 & 0.002 & -0.023 & 0.007 & 0.016& -0.032\\
 &  & 1& 0.711 & -0.025 & 0.041 & 0.144 &0.340\\
  &  &  & 1 & -0.064 & -0.154 & 0.065& 0.201\\
  &  &  &  & 1 & -0.032 & -0.022& -0.252\\
  &  &  &  &  & 1 & -0.017 &0.013\\
  &  & 
 &  &  &  & 1 &    0.483\\
   &  &  &  &  &  & & 1    \\  \hline
   \end{tabular} \end{center}
       \caption{ \label{tab:2} 
    Results of the global fit in our default scenario.
All parameters are expressed in $\GeV$ at the appropriate power and all, except $m_c$, are in the kinetic scheme at $\mu=1\GeV$. The first and second rows give central values and uncertainties, the correlation matrix follows. }
\end{table}

As should be clear from the above discussion on higher orders in the OPE, 
the estimate of theoretical errors and of their  correlation is crucial. We follow the strategy of \cite{Gambino:2004qm,Gambino:2013rza} for theoretical uncertainties, updating it because of the new corrections that we include. In particular, we assign an irreducible uncertainty of 8 MeV to $m_{c,b}$,
and vary $\as(m_b)$ by $\pm0.018$, $\mupi$ and $\mug$ by $\pm7\%$, $\rd$ and $\rls$ by $\pm 30\%$.  
This implies a total theoretical uncertainty between 2.0\% and 2.6\% in the semileptonic width, depending on the scheme.
For the theory correlations we adopt scenario D of Ref.\,\cite{Gambino:2013rza}, {\it i.e.}\ we assume no  correlation between different {\it central} moments and  a  correlation 
between  the same moment measured at different $E_{cut}$, depending on the
proximity of the cuts and their magnitude.  In the extraction of $|V_{cb}|$ we use the latest isospin average  $\tau_B=1.579(5)$ps \footnote{Y. Amhis {\it et al.} (Heavy Flavor Averaging Group), arXiv:1207.1158 [hep-ex]  and online update at http://www.slac.stanford.edu/xorg/hfag.}.

In Table \ref{tab:2} we show the results of the fit and the correlation matrix among the fitted 
parameters. With respect to the default fit of Ref.\,\cite{Gambino:2013rza}, 
 $|V_{cb}|$ is reduced  by 0.5\%,  see Eq.\,(\ref{incl}), $m_b^{kin}$ is increased by about 10 MeV, $\mupi$ and $\rd$ are both shifted upward by  about 10\%.
 As the method and inputs are the same of Ref.\,\cite{Gambino:2013rza}, except for the value of $\tau_B$ which only reflects in a tiny
+0.1\%  shift in $|V_{cb}|$, the 
difference  can be  mostly attributed to the new corrections.  Because of smaller theoretical errors, the final uncertainties are slightly reduced.
The $\chi^2/d.o.f.$ is very good, about 0.4.  
\begin{figure}
\includegraphics[width=8cm]{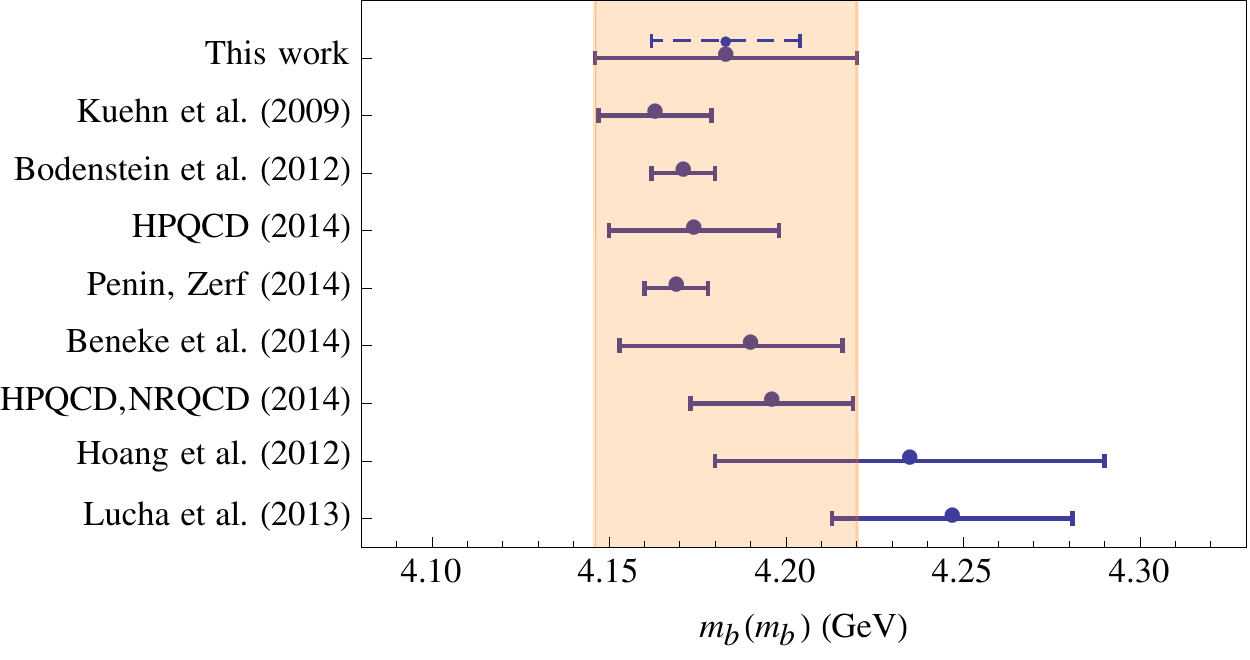}
\caption{ Comparison of different $\overline{m}_b(\overline{m}_b)$ determinations \cite{Bodenstein:2011fv, Beneke:2014pta,Colquhoun:2014ica,Chakraborty:2014aca,Hoang:2012us, Penin:2014zaa,Lucha:2013gta}. The dashed line denotes the error before scheme conversion.}
\label{mb}
\end{figure}

It is interesting to compare the $b$ mass extracted from the fit with other recent determinations, generally expressed in terms of $\overline{m}_b(\overline{m}_b)$   in the $\overline{\rm MS}$ scheme.
This is shown in Fig.\,\ref{mb}, after converting $m_b^{kin}$ into $\overline{m}_b(\overline{m}_b)$. The scheme conversion implies an additional $\sim 30$MeV uncertainty \cite{Gambino:2011cq}, enlarging the final error to 37\,MeV, because it is known only 
through $O(\as^2)$. Our result,
$\overline{m}_b(\overline{m}_b)=4.183(37)$GeV,
 agrees well with those reported in the Figure. The combination $m_b^{kin}(1{\rm GeV})\!-\!0.85 \overline{m}_c(3\GeV)$ is
best determined to  $3.714\pm0.018$GeV.

\begin{table}[t]
  \begin{center} \begin{tabular}{rcccccccc}\hline

 & $m_b^{kin}$ & $m_c$   &  $\mupi $ &$\rd$ &$\mug$ & $\rls$  & ${\rm BR}_{c\ell\nu}$ & $10^3|V_{cb}|$ \\ \hline

    $a)$& 4.561 & 1.092 & 0.464 & 0.175 & 0.333 & -0.146 & 10.66 & 42.04 \\  

    &0.021 & 0.020 & 0.067 & 0.040 & 0.061 & 0.096 & \ 0.16 &\ 0.67\\ \hline

      $b)$& 4.576 & 1.092 & 0.466 & 0.174 & 0.332 & -0.146 & 10.66 & 42.01 \\  

    &0.020 & 0.014 & 0.068 & 0.039 & 0.061 & 0.096 &\ 0.16 &\ 0.68\\ \hline

      $c)$& 4.548 & 0.985 & 0.467 & 0.168 & 0.321 & -0.146 & 10.66 & 42.31 \\  

    &0.017 & 0.012 & 0.068 & 0.038 & 0.058 & 0.096 & \ 0.16 & \ 0.76\\ \hline
   \end{tabular} \end{center}
       \caption{ \label{tab:3} 
    Results of the fit in different scenarios: $a)$ with 
 $m_c$ in the kinetic scheme, $m_c^{kin}=1.091(20)$GeV from \cite{Chetyrkin:2009fv};
 $b)$ in the $\overline{\rm MS}$ scheme at a lower scale, with $\overline{m}_c(2\rm GeV)=1.091(14)$GeV from \cite{Chetyrkin:2009fv}; $c)$ same as our default fit, with an additional constraint $m_b^{kin}=4.533(32)$GeV, derived from \cite{Chetyrkin:2009fv}. }
\end{table}

Table \ref{tab:3} shows the results  when the fit is performed with $m_c$ in a different scheme or at a different scale with respect to our default fit of Table \ref{tab:2}.
The results are remarkably consistent and very close to the default fit, with the only partial exception of $m_b$, which becomes 1$\sigma$ higher  when $\overline{m}_c(2\rm GeV)$ is used as input.
Table \ref{tab:3} also reports the results of a fit with an additional constraint on $m_b$. 
Even the currently most  precise $m_b$ determinations are spoiled by the uncertainty due to the scheme conversion to $m_b^{kin}$.
Because of this, and of the large range of $m_b$ values given in the literature,  we prefer to avoid using a $m_b$ constraint in our default fit.

\begin{figure}[t]
\includegraphics[width=8cm]{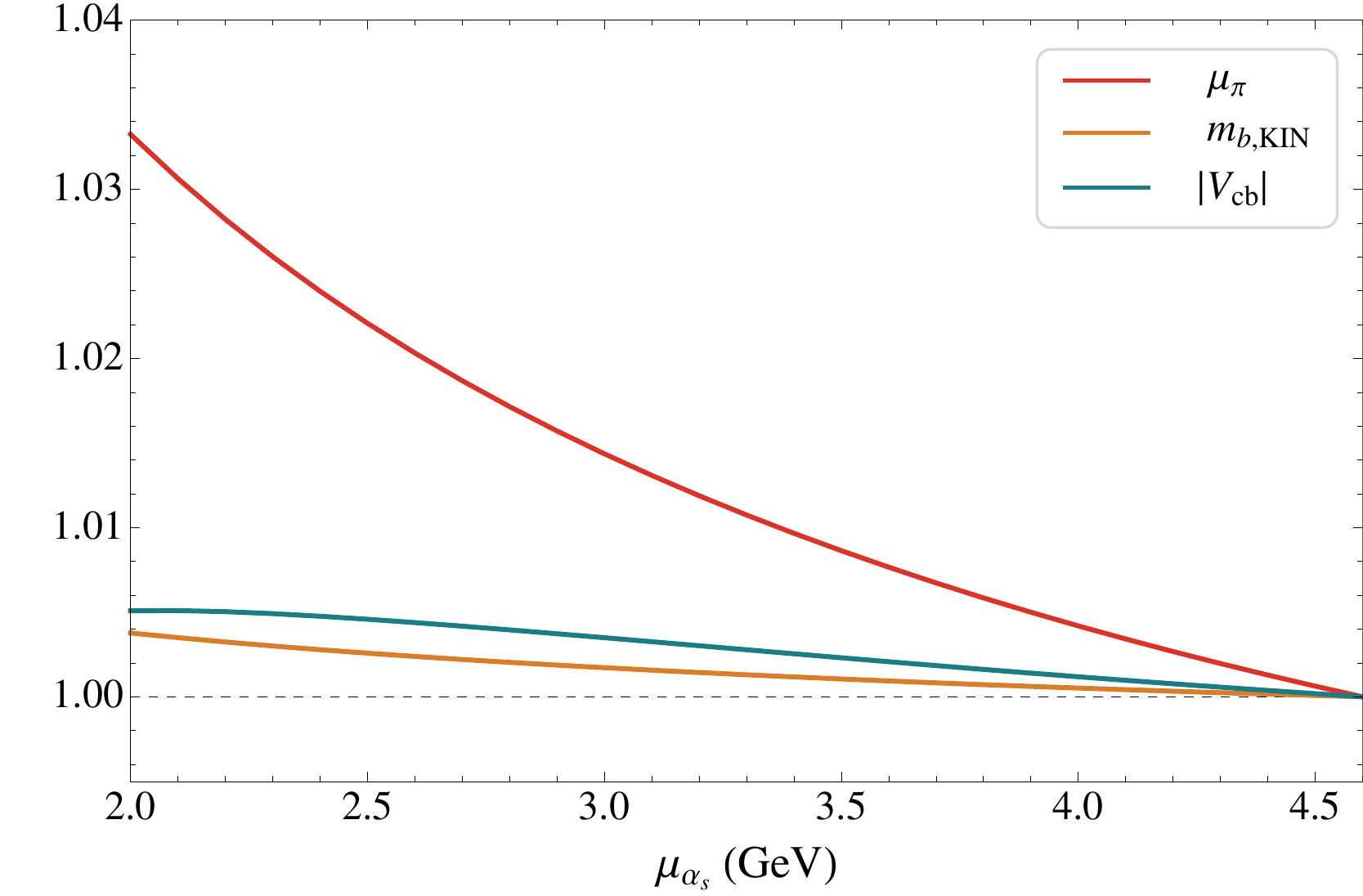}
\caption{ Relative variation of the central values for $|V_{cb}|$, $m_b^{kin}$, and $\mu_\pi^2$ on the scale of $\alpha_s$ in the default fit. }
\label{scaledep}
\end{figure}

\begin{figure}[t]
\includegraphics[width=8cm]{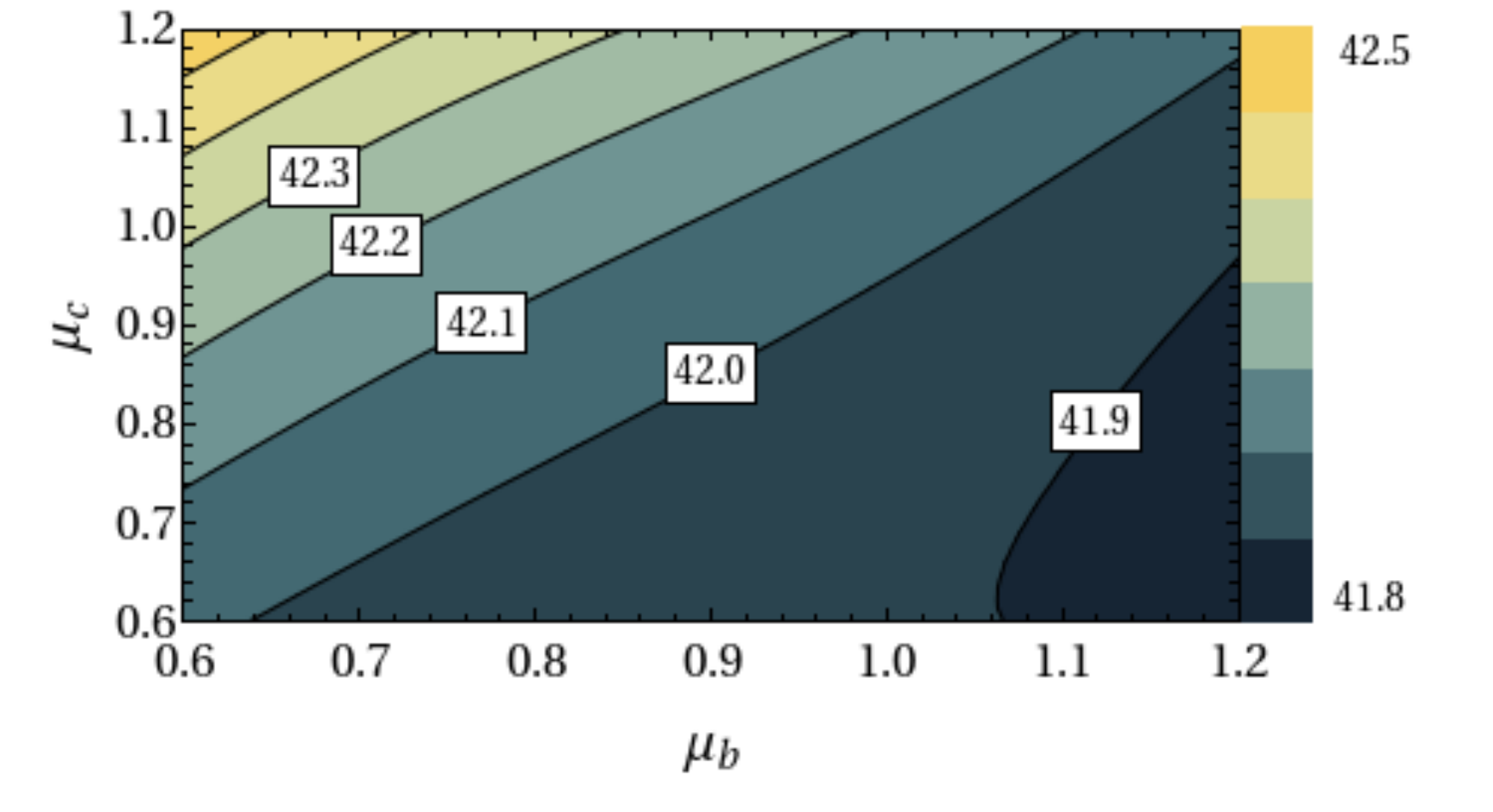}
\caption{ Dependence of the $|V_{cb}|$ central value in fit $a)$ on the kinetic
cutoff of the $b$ and $c$ masses. }
\label{scaledep2}
\end{figure}

  Overall, the fit results depend little on the scale of  $\as$. This is shown in Fig.\,\ref{scaledep} for the default fit. $|V_{cb}|$ and $m_b^{kin}$ increase by less than 0.5\% if we perform the whole analysis using   $\as(m_b/2)$, while $\mupi$ and in general the OPE parameters are slightly more sensitive. A similar behavior is observed for the fits in Table \ref{tab:3}. 
Fig.\,\ref{scaledep2} shows instead the $\mu_{kin}$ dependence of $|V_{cb}|$ in the case $a)$, keeping the scales of $m_b$ and $m_c$ distinct. In all cases,
 the scheme and scale dependence confirms the size of 
theoretical errors employed in our analysis. 

Finally, we  update the value of the semileptonic phase space ratio $C$, 
\[
C=\left| \frac{V_{ub}}{V_{cb}} \right|^2 
\frac{\Gamma[\bar{B} \to X_c e \bar{\nu}]}{\Gamma[\bar{B} \to X_u e \bar{\nu}]},
\]
which is
often used in the calculation of the branching ratio of radiative and rare semileptonic  $B$ decays, see \cite{Gambino:2013rza} for details. 
Using the default fit and $\mu_{WA}=m_b/2$,
 we find $C=0.574\pm 0.008\pm 0.014$,
where the first uncertainty comes from the parameters determined in the fit, and the second 
from unknown higher orders, estimated as explained above. Since the ratio $C$ receives large perturbative corrections when it is expressed in terms of $\overline{m}_c(3\rm GeV)$ \cite{Gambino:2013rza},  we believe that  using $\overline{m}_c(2\rm GeV)$ leads to a more reliable
estimate. Including the $m_b^{kin}$ mass constraint derived from \cite{Chetyrkin:2009fv}  as well, we find
\be
C=0.568\pm 0.007\pm 0.010,
\ee
slightly higher but with a smaller error than the corresponding value in \cite{Gambino:2013rza}.

\section{conclusion}
In summary, we have improved the inclusive determination of $|V_{cb}|$ 
through the inclusion of the complete $O(\as \Lambda_{\rm QCD}^2/m_b^2)$ effects.
Our final value,
\be
|V_{cb}|= (42.21\pm 0.78)\times 10^{-3},
\label{final}\ee
is compatible with previous analyses, but  its uncertainty is slightly reduced   thanks to the smaller theoretical errors.  Eq.\,(\ref{final})  still differs at the 2.9$\sigma$ level from Eq.\,(\ref{excl}). We find no sign of inconsistency in the inclusive analysis and adopt a conservative estimate of theory errors. The latter
could be further reduced by a calculation of $O(\as \Lambda^3_{\rm QCD}/m_b^3)$ contributions, as well as by a better understanding of higher power corrections, see \cite{Heinonen:2014dxa}.

\vspace{3mm}

\begin{acknowledgments}
We are grateful to M.~Misiak and C.~Schwanda for useful correspondence.
This work is supported in part by MIUR under contract 2010YJ2NYW 006,  by the EU Commission through the HiggsTools Initial Training Network PITN-GA-2012-316704,   and by Compagnia di San Paolo under contract ORTO11TPXK.
\end{acknowledgments}

\bibliography{draft}

\end{document}